\documentclass[preprint, superscriptaddress, showpacs,
aps]{revtex4}
\usepackage{graphicx}
\begin{document}

\title{ Probe the 2$s_{1/2}$ and 1$d_{3/2}$ state level inversion with electron-nucleus scattering}
\author{Wang Zai-jun}
\email{ zaijunwang99@hotmail.com} \affiliation{School of Science, Tianjin University of Technology and Education, Tianjin
300222, China}
\author{Ren Zhong-zhou}
\affiliation{Department of Physics, Nanjing University, Nanjing
210008, China } \affiliation{Center of Theoretical Nuclear
Physics, National Laboratory of Heavy-Ion Accelerator at Lanzhou,
Lanzhou 730000, China}
\author{Dong Tie-kuang} \affiliation{Purple Mountain Observatory, Chinese Academy of Sciences, Nanjing 210008, China }

\noindent
\begin{abstract}
The neutron-rich even-even nuclei $^{26-40}$Mg, $^{28-46}$Si,
$^{30-48}$S, and $^{32-56}$Ar are calculated with the RMF model
and the phase-shift electron scattering method. Results show that
level inversion of the 2$s_{1/2}$ and 1$d_{3/2}$ proton states may
occur for the magnesium, silicon, sulphur, and argon isotopes with
more neutrons away from the stability line. Calculations show that
the variation of the central charge densities for $^{30-48}$S, and
$^{32-56}$Ar are very sensitive to the 2$s_{1/2}$ and 1$d_{3/2}$
proton state level inversion, and the level inversion can lead to
a large measurable central charge depletion to the charge density
distributions for the neutron-rich isotopes. Calculations also
show that the charge density differences between the isotopes with
and without central charge depletion can reveal not only the level
inversion of the 2$s_{1/2}$ and 1$d_{3/2}$ proton states but also
the behavior of the proton wave functions of both states. The
results can provide references for the possible study of the
nuclear level inversion and nuclear bubble phenomenon with
electron scattering off short-lived nuclei at RIKEN or/and GSI in
the future. In addition, direct nuclear reaction $^{44}$S(n,
d)$^{43}$P or $^{44}$S($^{3}$H, $\alpha$)$^{43}$P might also be a
possible way to study the 2$s_{1/2}$ and 1$d_{3/2}$ proton state
level inversion.


\end{abstract}

\maketitle

\section{Introduction}

One of the hotly debated problem in nuclear physics is the
existence of the nuclear level inversion and the related nuclear
phenomena in exotic nuclei \cite{wim1,www1}. The problem of
nuclear level inversion has been studied for many years both
theoretically and experimentally and great progress has been
achieved \cite{Ots1,Ots2,Ots3,Uts,Pov,Cau}. However, there are
still some problems to which the solutions are not really clear.
For instance, the real causes for the nuclear level inversion are
not completely known yet and some results given by different
models or experiments do not agree with each other. Another
unsolved problem, the nuclear bubble phenomenon, is also related
to the nuclear level inversion. The nuclear bubble phenomenon has
also been studied for many years with a variety of nuclear models
\cite{wil46,sie67,Swi83,Dec03}. The majority opinion is that one
of the main causes for the nuclear bubble formation is the s-d
level inversion. However, we haven't detected any nuclear bubbles
yet. Do the nuclear bubbles really exist? If so, is the main cause
the s-d level inversion? The clarification of these problems
pleads for further elaborate experimental investigation on the
level inversion and the related nuclear phenomena. In terms of the
modern quantum theory, level inversion will lead to the change of
the nucleon distribution, so a very good method to probe the level
inversion could be to measure the nucleon distributions in exotic
nuclei. Electron scattering has proven to be an excellent tool for
measuring the charge density distributions and proton
distributions of nuclei \cite{hof56,hof57,Don84,sic01}. Therefore,
in this paper we focus on exploring the feasibility of studying
the s-d level inversion with electron-nucleus scattering
experiments.

In recent years, based on the development of the RI beam
technology, some new facilities for electron scattering off
short-lived nuclei have been constructed at different
laboratories. For instance, the double storage rings of MUSES
\cite{sud01,sud04,sud05,mot05} at RIKEN in Japan has made it
possible to perform the electron scattering experiments by storing
the radioactive ion beams in one ring and storing the electron
beams in another. A new novel internal target for electron
scattering on unstable nuclei, the SCRIT (self-confining
radioactive isotope ion target), has also been developed and
gained exciting success at RIKEN \cite{wak08}, and the first
demonstrative experiment of electron scattering off short-lived
nucleus $^{133}$Cs has been performed in 2009 \cite{sud001}. In
addition, a similar electron-ion collider at GSI in Germany
\cite{int02,sim07,rub06} has also been constructed. Very recently,
great progress in parity-violating electron-nucleus scattering has
been made at the Thomas Jefferson National Accelerator Facility in
the United States \cite{abr001}. In parallel with the development
of the new facilities and experimental researches, theoretical
studies on electron scattering off exotic nuclei have also
developed and some new results have been obtained
\cite{Gar,Ant,Sar,Kar,Zai1,Roc,Zai2,yan001,Ma001}. We believe that
the theoretical results will provide useful references for
experimental studies and that the newly built facilities will
provide good opportunities for further studying the nuclear level
inversion and the nuclear bubble problems with electron-nucleus
scattering experiments in the future.

The possible appropriate candidate nuclei for studying the s-d
level inversion and the proton bubble phenomenon could be the
isotopes $^{30-48}$S and $^{32-56}$Ar, since their outmost protons
just move in the 2$s_{1/2}$ or/and 1$d_{3/2}$ states, and this
will make the proton distributions of these nuclei more sensitive
to the s-d level inversion than those of the others. Along with
the sulfur and argon isotopes the magnesium and silicon isotopes
$^{26-40}$Mg and $^{28-46}$Si are also chosen for discussion and
comparison. We will calculate the variations of the 2s$_{1/2}$ and
1d$_{3/2}$ proton state energy levels and the variation of the
level gap between the two states with respect to the change of the
neutron number to find the possible 2$s_{1/2}$ and 1$d_{3/2}$
state level inversion. We also calculate the charge density
distributions, the elastic electron scattering form factors and
cross sections to further investigate the influences of the
2$s_{1/2}$ and 1$d_{3/2}$ state level inversion so as to study if
the 2$s_{1/2}$ and 1$d_{3/2}$ state level inversion and the proton
bubbles can be detected with electron nucleus scattering.

The method that we use is the combination of the RMF nucleus
structure theory and the phase shift analysis method for
electron-nucleus scattering. The RMF theory is currently a widely
used model in the calculations of stable as well as unstable
nuclei
\cite{wal1,wal2,wal3,gam1,hor1,ma1,ren3,ren4,Sug1,Jia1,tod1}.
Therefore, we use this theory to investigate the energy levels and
the proton occupation probabilities and produce the charge density
distributions. Phase shift analysis is a very stable method in
calculating the electron-nucleus scattering process in a wide
range of incident energies
\cite{Yenn001,Gar,Ant,Sar,Kar,Zai1,Roc,Zai2,yan001,Ma001}, so we
use this method to calculate the cross sections and form factors.

The paper is organized in the following way. Section II is a brief
review of the formalism of the phase shift analysis method for
elastic electron scattering. Section III is the numerical results
and discussions. A summary is given in Section IV.

\section{Formalism}
The elastic electron scattering process can be described by the
Dirac equation \cite{Rose}
\begin{equation}\label{Eq1}
   [{\alpha}\cdot\textbf{p}+\beta m +V(r)]\Psi(\textbf{r})=E\Psi(\textbf{r}),
\end{equation}
where $\alpha$ and $\beta$ are the Dirac matrices, $E$ and
$\textbf{p}$ are the energy and momentum of the incident
electrons, and $m$ is the rest mass of the electron. $V(r)$ is the
potential between the electron and the nucleus. To obtain the
differential cross section of the elastic electron scattering, we
must solve the above Dirac equation. In the following we introduce
the phase-shift analysis method. The details of this method can be
found in many quantum physics literatures
\cite{Yenn001,yan001,Ma001}, so we only give a brief review of it.

For a spherical scalar potential $V(r)$, the wave function of the
Dirac equation can be expanded in terms of a series of spherical
spinors with definite angular momenta \cite{Bjorken}
\begin{eqnarray}\label{Eq2}
    \Psi(\textbf{r})=\frac{1}{r}\left[\begin{array}{c}
                                    P(r)\Omega_{\kappa,m_j}(\theta,\phi) \\
                                    iQ(r)\Omega_{-\kappa,m_j}(\theta,\phi)
                                  \end{array} \right],
\end{eqnarray}
where $P(r)$ is the upper-component radial wave function, $Q(r)$
is the lower-component one, and $\Omega$ are the spherical
spinors. The functions $P(r)$ and $Q(r)$ satisfy
\begin{eqnarray}\label{Eq3}
     \frac{dP}{dr}=-\frac{\kappa}{r}P(r)+[E-V(r)+2m]Q(r), \\
  \label{Eq4}  \frac{dQ}{dr}=-[E-V(r)]P(r)+\frac{\kappa}{r}Q(r).
\end{eqnarray}
After determining the asymptotic behavior of $rV(r)$, we can
express the upper and lower radial wave functions at large
distances as
\begin{eqnarray}\label{Eq5}
P(r) = F^{(u)}(r)\cos \delta +G^{(u)}(r) \sin \delta,\\
\label{Eq6} Q(r) = F^{(l)}(r)\cos \delta +G^{(l)}(r) \sin \delta,
\end{eqnarray}
where $F^{(u,l)}$ and $G^{(u,l)}$ are the regular and irregular
Dirac spherical Coulomb functions. The symbols $u$ and $l$ stand
for the upper and lower components and $\delta$ is the phase
shift.

By solving the coupled radial equations Eqs. (\ref{Eq3}) and
(\ref{Eq4}) with the asymptotic conditions defined by Eqs.
(\ref{Eq5}) and (\ref{Eq6}), we can obtain the spin-up
($\delta_l^+$) and spin-down ($\delta_l^-$) phase shifts for the
partial wave with orbital angular momentum $l$. Then we can
determine the direct scattering amplitude by
\begin{eqnarray}\label{Eq7}
    f(\theta)=\frac{1}{2ik}\sum_{l=0}^{\infty}(l+1)(e^{2i\delta_l^+}-1)P_l(\cos\theta)
    \nonumber\\
     +l(e^{2i\delta_l^-}-1)P_l(\cos\theta),
\end{eqnarray}
and the spin-flip scattering amplitude by
\begin{eqnarray}\label{Eq8}
    g(\theta)=\frac{1}{2ik}\sum_{l=0}^{\infty}\left[e^{2i\delta_l^-}-e^{2i\delta_l^+}\right]P_l^1(\cos\theta)
    \,,
\end{eqnarray}
where $P_l$ and $P_l^1$ are the Legendre polynomials and
associated Legendre functions, respectively. The differential
cross section for the elastic electron-nucleus scattering can be
obtained as follows
\begin{eqnarray}\label{Eq9}
    \frac{d\sigma}{d\Omega}=|f(\theta)|^2+|g(\theta)|^2 \, .
\end{eqnarray}

After the differential cross sections are obtained, the charge
form factors squared $|F(q)|^2$ can be calculated by dividing the
differential cross sections with the Mott cross section
$({d\sigma}/{d\Omega})_M$
\begin{eqnarray}\label{Eq10}
|F(q)|^2=\frac{{d\sigma}/{d\Omega}}{({d\sigma}/{d\Omega})_M},
\end{eqnarray}
where
\begin{eqnarray}\label{Eq11}
(\frac{d\sigma}{d\Omega})_M=\frac{\alpha^2(\hbar
c)^2\cos^2\frac{\theta}{2}}
 {4E^2\sin^4\frac{\theta}{2}}.
\end{eqnarray}

For high-energy electron scattering off light nuclei, the recoil
of the target nucleus must be taken into account. We do this by
dividing the differential cross sections by the factor \cite{Sic2}
\begin{eqnarray}\label{Eq12}
f_{\mathrm{rec}} =
\left(1+\frac{2E\sin^{2}\frac{\theta}{2}}{Mc^{2}}\right),
\end{eqnarray}
where $M$ is the mass of the nucleus and $E$ is the energy of the
incident electrons. Another correction that should be considered
is the attraction felt by the electrons, although this effect is
not very strong for light nuclei. We do this with the standard
method in electron scattering, that is to replace the momentum
transfer $q$ with the effective momentum transfer
\begin{eqnarray}\label{Eq13}
q_{\mathrm{eff}} = q\left(1 + \frac{3}{2}\frac{��Z\hbar
c}{ER_0}\right),
\end{eqnarray}
in our calculation, where $R_0$ = 1.07A$^{1/3}$ and $A$ is the
mass number of the nucleus.

In numerical calculations, we take $V(r)$ as
\begin{eqnarray}\label{Eq14}
V(r) =e\frac{1}{r}{\int_{0}}^{r}\rho_{ch}(r^{'})4\pi
{r^{'}}^2dr^{'}
 +e{\int_{r}}^{\infty} \rho_{ch}(r^{'})4\pi {r^{'}}dr^{'},
\end{eqnarray}
where $\rho_{ch}(r')$ is the charge densities that are obtained by
folding the point proton densities, which are calculated from the
RMF model, with the proton charge density distribution \cite{gre1}
\begin{eqnarray}\label{Eq15}
\rho_p(r) = \frac{Q^3}{8\pi}e^{-Qr},
\end{eqnarray}
where $Q^2$ = 18.29 fm$^2$ = 0.71 GeV$^2$ ($\hbar c $ = 0.197 GeV
fm = 1). The corresponding rms charge radius of the proton is
$r_p$ = 0.81 fm.

The center-of-mass effect is also taken into account in the
calculation. The correction to the binding energy due to the
center-of-mass fluctuations is included by subtracting $30.75 \
\mathrm{MeV} A^{-1/3}$ \cite{Rein1}, a non-relativistic estimation
of the kinetic energy of the center-of-mass, from the total
energy. The center-of-mass fluctuations also have effect on the
charge form factors. This effect is taken into account by
multiplying the form factors with the following factor
\cite{Rein2}
\begin{eqnarray}\label{Eq16}
f_{CM}(q)=\mathrm{exp}\left(\frac{q^{2}}{8\langle
\hat{P}^{2}_{CM}\rangle}\right),
\end{eqnarray}
where $\langle\hat{P}^{2}_{CM}\rangle$ can be estimated from the
kinetic energy of the center-of-mass.

The RMF model has been developed into a standard nuclear structure
theory and has been extensively used to describe the properties of
the ground and low excited states both for stable and unstable
nuclei. The details for the RMF model can be found in many
articles such as
\cite{wal1,wal2,wal3,gam1,hor1,ma1,ren3,ren4,Sug1,Jia1,tod1}. We
will not redundantly depict them here.

\section{Numerical results and discussions}
We first give the RMF model results. With the NL-SH parameter set,
we calculated the single nucleon state energy levels and their
occupation probabilities. The numerical results for the
1$d_{5/2}$, 2$s_{1/2}$ and 1$d_{3/2}$ proton states which are
involved in our discussions are listed in Table 1. By careful
study, it is found from Table 1 that the energy levels and
occupation probabilities have two features. The first one is that
for each nuclei the energy level of the 1$d_{5/2}$ state is much
lower than those of the 2$s_{1/2}$ and 1$d_{3/2}$ states, and this
consequently leads to a much greater proton occupation probability
of the 1$d_{5/2}$ state than those of the 2$s_{1/2}$ and
1$d_{3/2}$ states. For $^{26-40}$Mg, the 1$d_{5/2}$ state
occupation probabilities are approximately 0.65, so the proton
occupation number in 1$d_{5/2}$ state is nearly 4. Whereas for the
2$s_{1/2}$ and 1$d_{3/2}$ states, the occupation probabilities are
less than 0.03, which corresponds to a proton occupation number of
less than 0.18. Thus, for the ground states of $^{26-40}$Mg the
outmost 4 protons will move in the 1$d_{5/2}$ state. For
$^{28-46}$Si, the 1$d_{5/2}$ state occupation probabilities are
approximately 0.95, which corresponds to nearly 6 protons. Whereas
for the 2$s_{1/2}$ and the 1$d_{3/2}$ state, the occupation
probabilities are less than 0.09 and 0.06, respectively, and the
occupation number of protons are also very small, just 0.18 in
2$s_{1/2}$ state and 0.24 in 1$d_{3/2}$. Therefore, the outmost 6
protons in $^{28-46}$Si move only in the 1$d_{5/2}$ state. The
occupation probabilities are too small for protons to stay in the
2$s_{1/2}$ and 1$d_{3/2}$ states. For $^{30-48}$S and
$^{32-56}$Ar, by the similar analysis, we can also derive that the
1$d_{5/2}$ state are all completely filled. Nevertheless, the
2$s_{1/2}$ and 1$d_{3/2}$ states each is not completely empty both
because the total number of protons are more than the lower states
can accommodate and because both states each has a relatively
large occupation probability.
\begin{table}[t]
\scriptsize \caption{\scriptsize The 1$d_{5/2}$, $2s_{1/2}$ and
$1d_{3/2}$ proton state energy levels and proton occupation
probabilities for $^{26-40}$Mg, $^{28-46}$Si, $^{30-48}$S, and
$^{32-56}$Ar. }
\begin{tabular*}{160mm}{c@{\extracolsep{\fill}}|ccccccc}\hline\hline
             &nuclide&$\epsilon({1d_{5/2}})$& $ \epsilon({2s_{1/2}})$  & $ \epsilon({1d_{3/2}})$&$p({1d_{5/2}})$& $ p({2s_{1/2}})$&$p({1d_{3/2}})$\\ \hline
Mg           &$^{26}$Mg&-11.165      &-4.022   & -2.526&0.65061& 0.02669 &0.01807\\
             &$^{28}$Mg&-13.411      &-6.518   & -5.035&0.65068 & 0.02662 &0.01785\\
             &$^{30}$Mg&-15.668      &-8.783   & -7.744&0.65064 & 0.02487 &0.01864\\
             &$^{32}$Mg&-17.881      &-10.838  & -10.351&0.65089& 0.02223 &0.01938\\
             &$^{34}$Mg&-19.976      &-12.704  & -12.715&0.65140 & 0.01957 &0.01963\\
             &$^{36}$Mg&-22.005      &-14.488  & -15.013&0.65172 & 0.01724 &0.02002\\
             &$^{38}$Mg&-23.971      &-16.206  & -17.258&0.65183 & 0.01525 &0.02059\\
             &$^{40}$Mg&-25.880      &-17.867  & -19.444&0.65178 & 0.01356 &0.02130\\  \hline
Si           &$^{28}$Si&-10.838      &-3.513   & -2.072&0.94509 & 0.08339 &0.04526 \\
             &$^{30}$Si&-13.169     &-5.899   & -4.658&0.94651 & 0.07822 &0.04563\\
             &$^{32}$Si&-15.447      &-8.138   & -7.350&0.94716 & 0.06952 &0.04894\\
             &$^{34}$Si&-17.650      &-10.269  & -9.914&0.94776 & 0.0612  &0.05212 \\
             &$^{36}$Si&-19.816      &-12.141  & -12.315&0.94964 & 0.05038 &0.05450 \\
             &$^{38}$Si&-21.569      &-13.719  & -14.265&0.95087 & 0.04362 &0.05587 \\
             &$^{40}$Si&-23.466      &-15.375  & -16.393&0.95162 & 0.03670 &0.05807 \\
             &$^{42}$Si&-25.027      &-16.848  & -18.139&0.95208 & 0.03295 &0.05913\\
             &$^{44}$Si&-25.994      &-18.028  & -19.182&0.95279 & 0.03373 &0.05757 \\
             &$^{46}$Si&-26.740      &-19.034  & -19.958&0.95369 & 0.03558 &0.05519\\  \hline
S            &$^{30}$S&-10.337       &-3.288  & -2.069&0.98090 & 0.53283 &0.2657\\
             &$^{32}$S&-12.624       &-5.551  & -4.542&0.98162 & 0.50663 &0.27762\\
             &$^{34}$S&-14.898       &-7.633  & -7.177&0.98203 & 0.42434 &0.31809\\
             &$^{36}$S&-17.049       &-9.690  & -9.622&0.98231 & 0.36378 &0.34789\\
             &$^{38}$S&-19.155       &-11.448  & -11.933&0.98310 & 0.2785 &0.3892\\
             &$^{40}$S&-21.242       &-13.105  & -14.242&0.98365 & 0.19231 &0.43134\\
             &$^{42}$S&-23.246       &-14.701  & -16.476&0.98388 & 0.12962 &0.46225\\
             &$^{44}$S&-24.948       &-16.174  & -18.362&0.98398 & 0.09915 &0.47724\\
             &$^{46}$S&-25.818       &-17.392  & -19.251&0.98442 & 0.11707 &0.46755\\
             &$^{48}$S&-26.520       &-18.478  & -19.936&0.98490 & 0.14663 &0.45197\\  \hline
Ar           &$^{32}$Ar&-9.862      &-3.055   & -2.189&0.98605  & 0.78843 &0.62985\\
             &$^{34}$Ar&-12.158      &-5.129   & -4.617&0.98662  & 0.74984 &0.6482\\
             &$^{36}$Ar&-14.406      &-7.069   & -7.130&0.98700  & 0.67273 &0.68614\\
             &$^{38}$Ar&-16.489      &-9.167   & -9.392&0.98722  & 0.64734 &0.69845\\
             &$^{40}$Ar&-18.604      &-10.840  & -11.659&0.98802 & 0.55036 &0.7456\\
             &$^{42}$Ar&-20.700      &-12.390  & -13.938&0.98893 & 0.43304 &0.80272\\
             &$^{44}$Ar&-22.695      &-13.883  & -16.144&0.98977 & 0.33355 &0.85107\\
             &$^{46}$Ar&-24.505      &-15.318  & -18.143&0.99044 & 0.26838 &0.88252\\
             &$^{48}$Ar&-25.455      &-16.679  & -19.175&0.99021 & 0.29548 &0.8693\\
             &$^{50}$Ar&-26.251      &-17.986  & -20.039&0.98984 & 0.34176 &0.84674\\
             &$^{52}$Ar&-27.121      &-19.219  & -21.085&0.98949 & 0.36164 &0.83733\\
             &$^{54}$Ar&-28.089      &-20.447  & -22.354&0.98907 & 0.35210 &0.84277\\
             &$^{56}$Ar&-29.091&-21.657  & -23.687&0.98864 & 0.33208 &0.85348\\
             \hline\hline
\end{tabular*}
\end{table}

The second feature is that for the isotopes with relatively
smaller neutron numbers the energy level of the 2$s_{1/2}$ state
is lower than that of the 1$d_{3/2}$ state, and with the increase
of the neutron number the 2$s_{1/2}$ and 1$d_{3/2}$ state levels
both decrease, however the 1$d_{3/2}$ state energy lowers more
rapidly than that of the 2$s_{1/2}$ state and, as a result, this
leads to the level inversion of the 2$s_{1/2}$ and 1$d_{3/2}$
states for some neutron-rich isotopes away from the stability
line. In order to show more clearly the trend of variation of the
energy levels of the 2$s_{1/2}$ and 1$d_{3/2}$ states
\begin{figure}[htb]
\includegraphics[width=7cm]{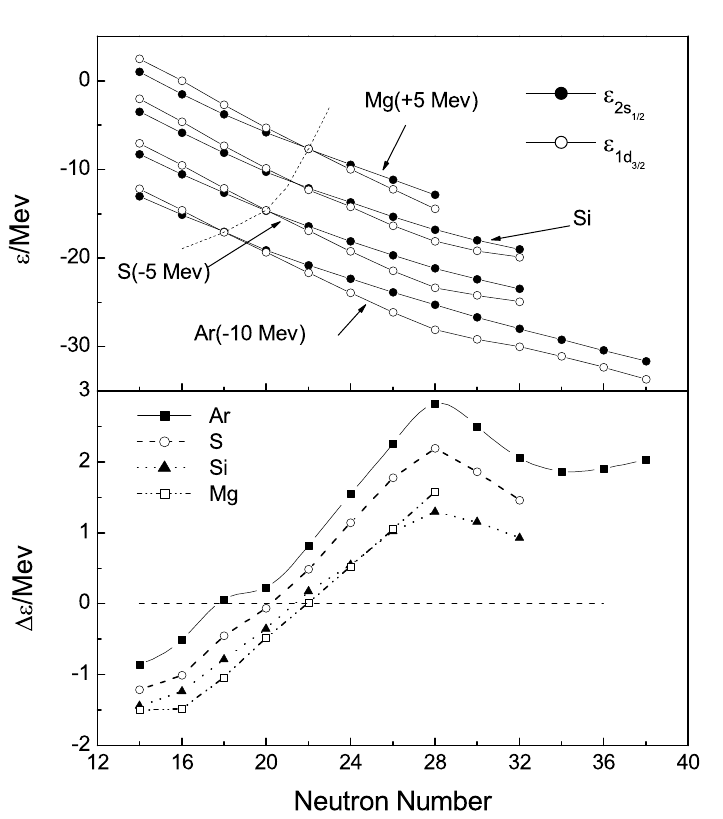}
\vspace{0cm}\caption{The upper panel is the 2$s_{1/2}$ and
1$d_{3/2}$ proton state levels of $^{26-40}$Mg, $^{28-46}$Si,
$^{30-48}$S, and $^{32-56}$Ar. For clearness, the curves for
$^{32-56}$Ar and $^{30-48}$S are shifted down 10Mev and 5Mev
respectively, and $^{26-40}$Mg up 5Mev. The lower panel is the
level gap between the 2$s_{1/2}$ and 1$d_{3/2}$ proton states.}
\end{figure}
\begin{figure}[htb]
\includegraphics[width=7cm]{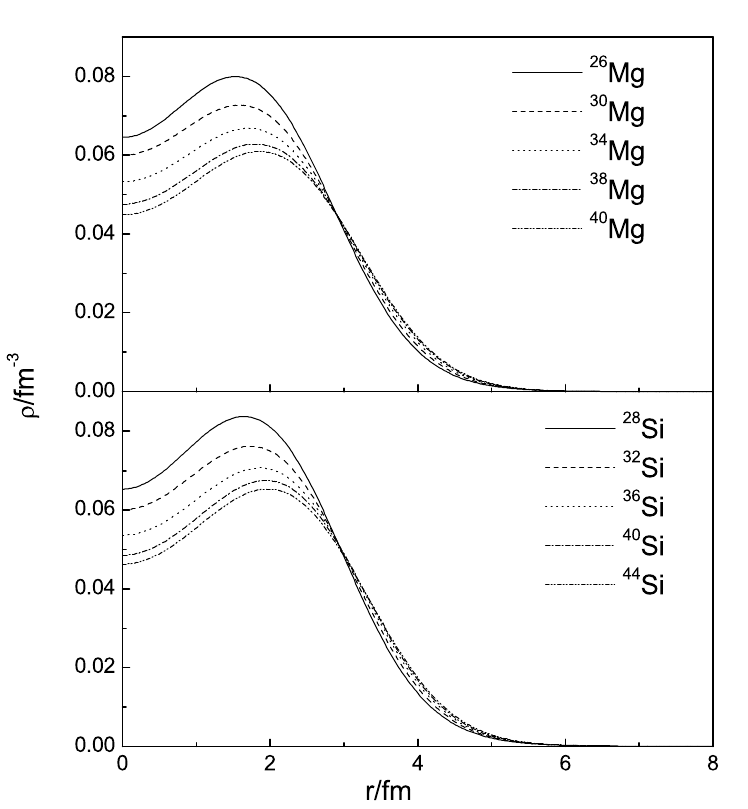}
\vspace{0cm}\caption{The charge density distributions of the
selected magnesium, silicon isotopes.}
\end{figure}
with respect to the increase of the neutron number, we have
plotted the 2$s_{1/2}$ and the 1$d_{3/2}$ state energy levels
$\varepsilon_{2s_{1/2}}$ and $\varepsilon_{1d_{3/2}}$ and the
level gap
$\triangle\varepsilon=\varepsilon_{2s_{1/2}}-\varepsilon_{1d_{3/2}}$
in Figure 1. The upper panel of Figure 1 shows that for each of
the four isotones with neutron number $N=14$ the energy level of
the 2$s_{1/2}$ state is lower than that of the 1$d_{3/2}$ state,
but as the neutron number increases, the 2$s_{1/2}$ and 1$d_{3/2}$
state energy levels both become lower and closer. As the neutron
number increases still more, the energy level curves cross each
other and the level inversion of the 2$s_{1/2}$ and 1$d_{3/2}$
states occurs. From the lower panel of Figure 1 we can see that
for magnesium, silicon and sulphur, the possible level inversion
occurs for the isotopes with $N>20$. For argon, the possible level
inversion occurs for the isotopes with $N\geq18$. In the following
paragraphs, it can be found that both features are very important
in accounting for the variation of the charge density
distributions of these nuclei.

In addition to the above two features, Figure 1 seems also to
reveal that the intersection of the 2$s_{1/2}$ and 1$d_{3/2}$
state level curves has the tendency to move towards the isotopes
with relatively smaller neutron numbers with the increase of the
proton number, as the dashed curves indicate in the upper and
lower panels of the figure. This may mean that for an isotonic
chain in the s-d shell region the level inversion of the
2$s_{1/2}$ and 1$d_{3/2}$ states occur more easily for the
proton-rich isotones. For instance, on the $N=18$ isotonic chain
the nucleus $^{36}$Ar may have the 2$s_{1/2}$ and 1$d_{3/2}$ state
level inversion, while $^{30}$Mg, $^{32}$Si, $^{34}$S may not.

In Figure 2 and 3 we present the charge density distributions
calculated by using the RMF model with the NL-SH parameter set.
For the sake of clearness, we did not give all the results of the
nuclei listed in Table 1 here. The charge density distribution
curves given in Figure 2 and 3 are typical of the charge
distribution shapes of the nuclei considered. It can be found from
Figure 2 that the charge density distributions of the five
magnesium isotopes each shows a noticeable depression around the
center, and the shapes of the charge distributions are very
similar. These features also hold true for the charge density
distributions of the silicon isotopes. While for sulphur and argon
isotopes, the results in Figure 3 show different features. For
sulphur, the isotopes $^{42}$S and $^{46}$S each has a significant
central charge density depletion, but $^{30}$S, $^{32}$S and
$^{34}$S only show a slight depression. The central depression
gets more and more noticeable as we go from $^{30}$S to $^{46}$S,
and the charge density distribution around the center varies from
nearly flattened to considerably depressed. For the argon isotopes
the similar conclusion can also be drown from Figure 3. Thus for
sulphur and argon, unlike magnesium and silicon, the charge
density distributions are not similar for each isotope. The shapes
of the charge distributions show an outstanding change around the
center as the neutron number increases.

The similarity of the charge density distributions between the
magnesium isotopes and between the silicon isotopes can be
explained in terms of the shell theory based on the results in
Table 1. In fact, the similarity of the charge density
distributions reveals that the isotopes do not have much
difference in proton configuration. This is consistent with the
results given in Table 1. As has been discussed in the previous
paragraphs, the proton configurations for the ground states of the
magnesium isotopes should be the same, i.e.
$(1s_{1/2})^{2}(1p_{3/2})^{4}(1p_{1/2})^{2}(1d_{5/2})^{4}$.
Although the RMF model results show that the level inversion of
the 2$s_{1/2}$ and the 1$d_{3/2}$ states may occur for the
magnesium isotopes with $N>20$, their proton configurations keep
unaffected, since the 2$s_{1/2}$ and 1$d_{3/2}$ states are empty
in the ground states. Since the shape of the wave function of a
proton is determined by its orbit, the same proton configurations
must correspond to the wave functions of the same shape. Thus, it
follows that the shapes of the proton wave functions are the same
for the magnesium isotopes.  Therefore, it is not curious that the
shapes of charge density distributions of the magnesium isotopes
are similar, since the charge density distribution of a nucleus is
determined by the wave functions of the protons. Likewise, the
similarity of the shapes of the charge density distributions for
the silicon isotopes can be explained.
\begin{figure}[htb]
\includegraphics[width=7cm]{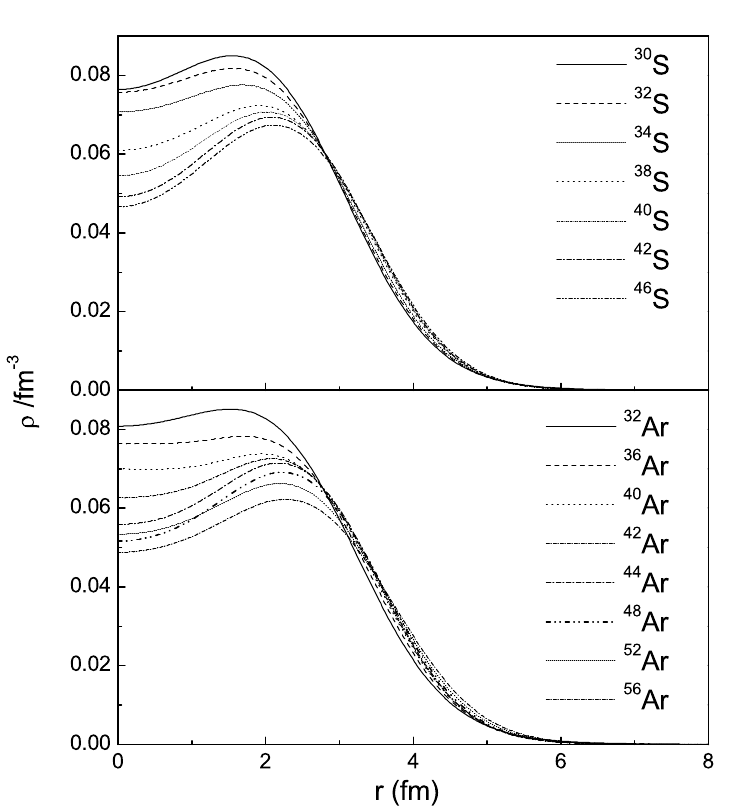}
\vspace{0cm}\caption{The charge density distributions of the
selected sulphur and argon isotopes.}
\end{figure}

Now we turn to the central depletion of the charge density
distributions. We know that the radial wave functions of the
$1s_{1/2}$, $1p_{3/2}$, $1p_{1/2}$ and $1d_{5/2}$ states have no
nodes, so the square of the wave functions each has only one peak.
For the $1s_{1/2}$ state wave function, the peak is always at the
center, but for the $1p_{3/2}$, $1p_{1/2}$ and $1d_{5/2}$ state
wave functions, the peaks are always away from the center. Since
the $1s_{1/2}$ state always has the lowest energy, the $1s_{1/2}$
state will not be empty for any nuclei in the ground states.
Therefore, the $1s_{1/2}$ state will guarantee that no nuclei in
the ground states can be
\begin{figure}[htb]
\includegraphics[width=7cm]{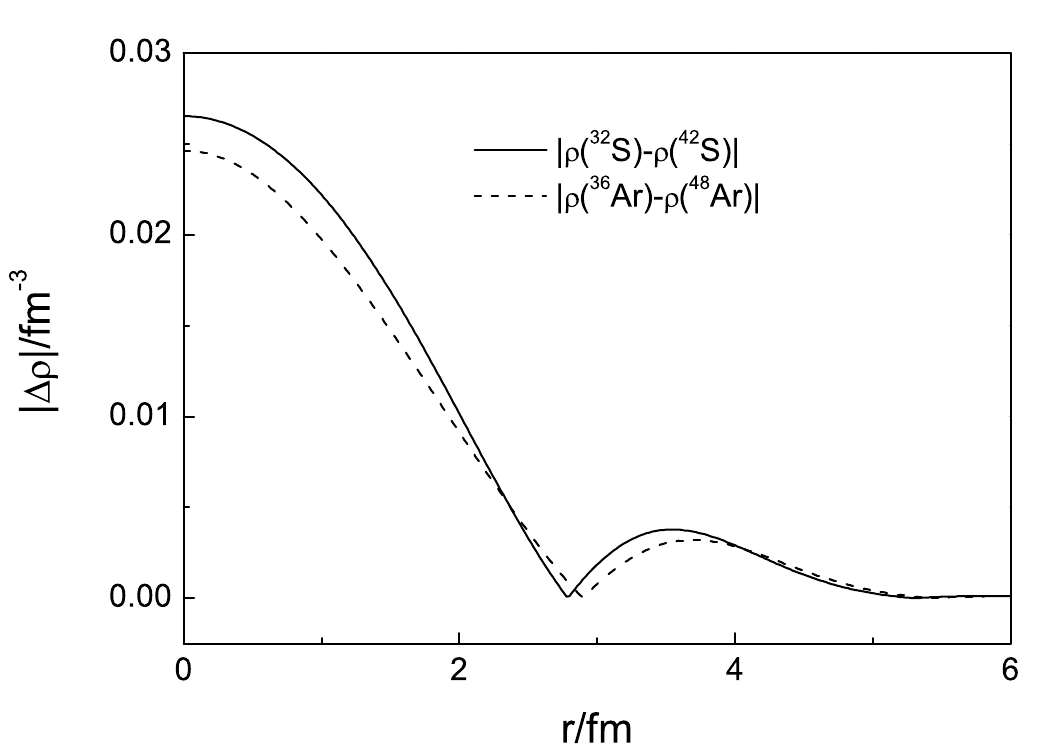}
\vspace{0cm}\caption{The charge density differences between
$^{32}$S and $^{42}$S and between $^{36}$Ar and $^{48}$Ar.}
\end{figure}
completely hollow in the central region. However, the nuclear
central charge depletion is still possible. On the one hand, this
is because that the number of the protons which can move in the
$1s_{1/2}$ state is limited, at most 2; and on the other hand,
this is because the peaks of the squared wave functions of the
$1p_{3/2}$, $1p_{1/2}$ and $1d_{5/2}$ states are away from the
center with up to 12 protons. If the peaks of the squared the wave
functions of the $1p_{3/2}$, $1p_{1/2}$ and $1d_{5/2}$ states
overlap or mostly overlap, the center depletion will probably
happen. The magnesium and silicon isotopes are this kind of nuclei
with the $1p_{3/2}$, $1p_{1/2}$ and $1d_{5/2}$ states nearly full
of protons and the peaks of their squared wave functions
overlapping in a large part, and this leads to the charge density
depletions at the center.

For sulphur and argon, as we have pointed out in the previous
paragraphs that with the increase of the neutron number the charge
density distributions of the isotopes vary outstandingly in the
central region. Why do the charge distribution shapes of these
isotopes around the center get depressed with an increasing
neutron number? One reason is that the level inversion of
2$s_{1/2}$ and 1$d_{3/2}$ states occurs as the neutron number
increases. As we have known that a particle prefers to move in a
lower energy state, so the energy level inversion of 2$s_{1/2}$
and 1$d_{3/2}$ states will greatly influence the occupation number
of the outmost protons in the two states. The effect can be
clearly seen in Table 1. For sulphur, from $^{30}$S to $^{48}$S,
the proton occupation number of the 2$s_{1/2}$ state decreases
from 1.07 to 0.20, while that of the 1$d_{3/2}$ state increases
from 1.06 to 1.91. The variations of the proton occupation numbers
of the 2$s_{1/2}$ and 1$d_{3/2}$ states for the argon isotopes are
similar. Another reason is that the squared wave function of the
2$s_{1/2}$ state has a main peak at the center, whereas the
squared wave function peak of the 1$d_{3/2}$ state is away from
the center. The occupation of 2$s_{1/2}$ state by protons will
enlarge the central charge densities, but the filling of the
1$d_{3/2}$ state by protons will increase the charge densities
away from the center. Thus, it is the combination of both the
causes that leads to the variations of the charge density
distributions for the sulphur and argon isotopes presented in
Figure 3.
\begin{figure}[htb]
\includegraphics[width=7cm]{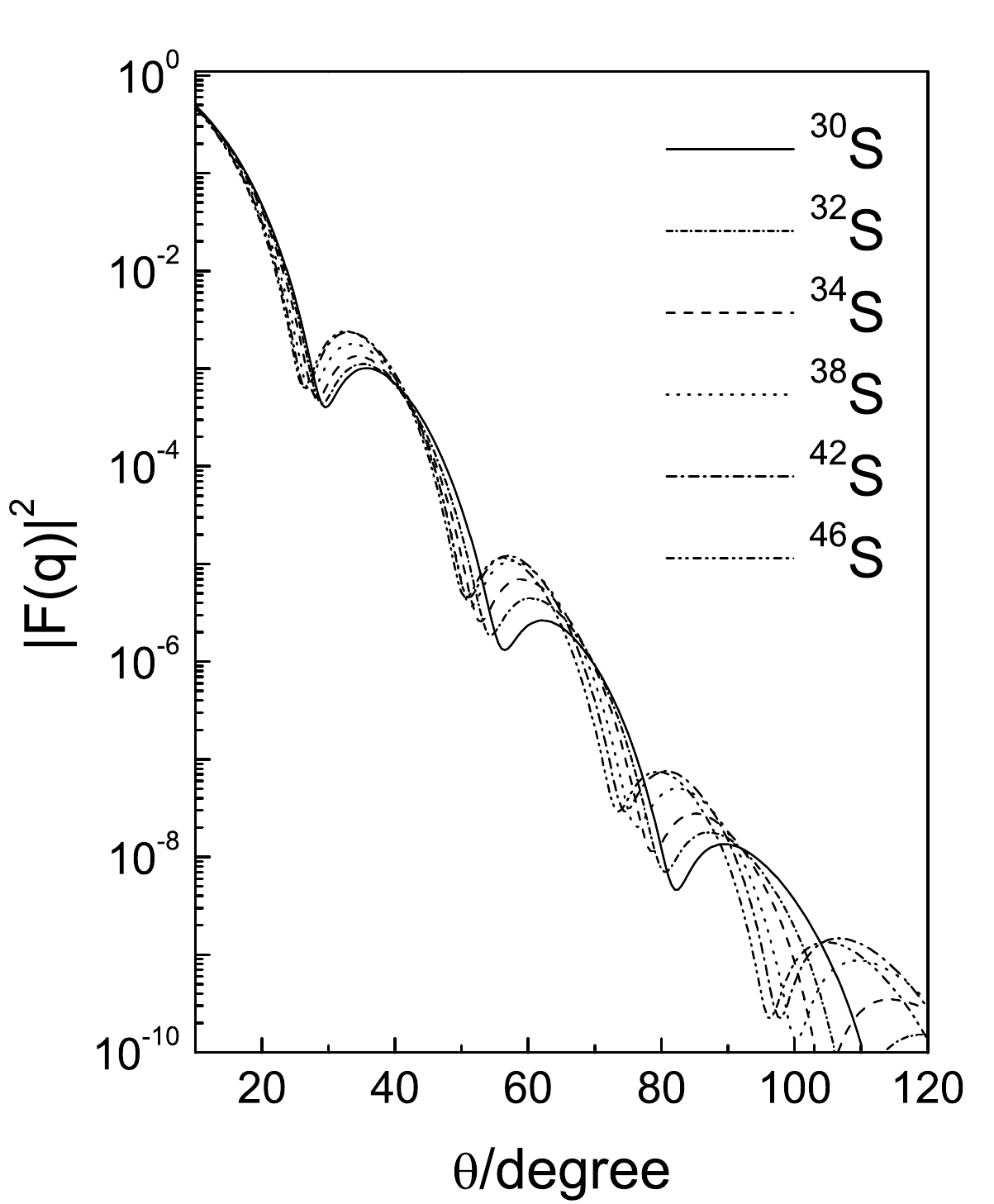}
\vspace{0cm}\caption{The charge form factors of
$^{30}$S,$^{32}$S,$^{34}$S,$^{38}$S,$^{42}$S and $^{46}$S
calculated with the phase shift analysis method.}
\end{figure}

The above discussion also shows that the variation of the central
charge densities, i.e. the depression of the charge densities, is
quite a good reveal of the combined effect of the 2$s_{1/2}$ and
1$d_{3/2}$ state level inversion and the behavior of the
2$s_{1/2}$ and the 1$d_{3/2}$ state wave functions at the center.
In Figure 4 we present two samples of the charge density
differences between two sulphur and between two argon isotopes.
The results show that the charge density differences between
$^{32}$S and $^{42}$S and between $^{36}$Ar and $^{48}$Ar are
large enough, especially near the center, to be observable
experimentally \cite{Cele1}. Therefore, it can be possible to
detect the 2$s_{1/2}$ and 1$d_{3/2}$ state level inversion and
study the behavior of the wave functions of 2$s_{1/2}$ and
1$d_{3/2}$ states experimentally by measuring the charge density
differences between the sulphur isotopes and between the argon
isotopes. A feasible and precise way of measuring nuclear charge
densities is elastic electron nucleus scattering \cite{hof56,
hof57, Don84, sic01}. Especially with the application of new
technology of measurement and the advent of the new-generation
electron-nucleus collider, not only has the measurement become
more precise, but the experiments of electron scattering off
short-lived nuclei have also become reality. Thus, we believe that
with the further development of new experimental technologies in
the future it could be possible to study the s-d level inversion
by measuring the charge density distributions of the exotic nuclei
with elastic electron nucleus scattering experiments.
\begin{figure}[htb]
\includegraphics[width=7cm]{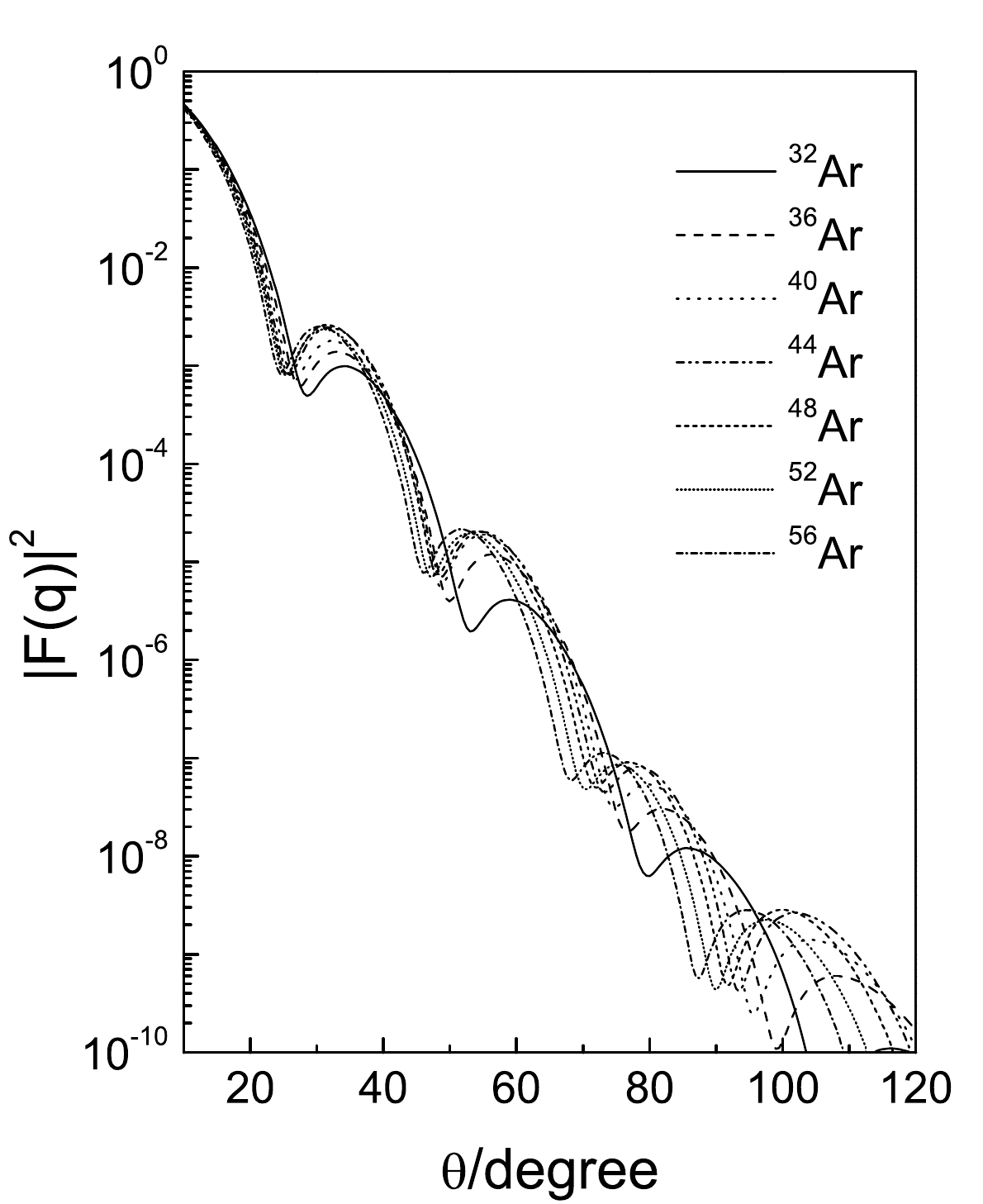}
\vspace{0cm}\caption{The charge form factors of $^{32}$Ar,
$^{36}$Ar, $^{40}$Ar, $^{44}$Ar, $^{48}$Ar, $^{52}$Ar and
$^{56}$Ar calculated with the phase shift analysis method.}
\end{figure}

To provide useful references for experimental research and
comparison between theoretical results and experimental data, we
give the phase shift analysis calculations of the charge form
factors for some candidate sulphur and argon isotopes. Figure 5 is
the theoretical charge form factors for the sulphur isotopes
$^{30}$S, $^{32}$S, $^{34}$S, $^{38}$S, $^{42}$S and $^{46}$S, and
Figure 6 gives the theoretical charge form factors for the argon
isotopes $^{32}$Ar, $^{36}$Ar, $^{40}$Ar, $^{44}$Ar, $^{48}$Ar,
$^{52}$Ar and $^{56}$Ar. It can be seen from both figures that the
charge form factors shift upward and inward noticeably as the
neutron number increases, and the largest shifts appear near the
maximums and minimums. Above all, the shifts of the form factors
between the isotopes with and without central charge density
depression, for instance those between $^{42}$S and $^{32}$S, are
large enough and can possibly be measured with elastic
electron-nucleus scattering on the new generation electron-nucleus
collider, and hence the charge density differences can be
extracted. In addition to the charge form factors, we further
calculated the differential cross section differences $D(\theta)$,
where
\begin{eqnarray} D(\theta) =
\frac{({d\sigma}(\theta)/{d\Omega})_{1}-({d\sigma}(\theta)/{d\Omega})_{2}}
{({d\sigma}(\theta)/{d\Omega})_{1}+({d\sigma}(\theta)/{d\Omega})_{2}}.
\end{eqnarray}
The upper panel of Figure 7 shows the differential cross section
differences between $^{34,38,42}$S and their stable isotope
$^{32}$S, and the lower panel between $^{40,48,56}$Ar and their
stable isotope $^{36}$Ar. Because of the 2$s_{1/2}$ and 1$d_{3/2}$
state level inversion, the charge density distributions of
$^{38,42}$S
\begin{figure}[htb]
\includegraphics[width=7cm]{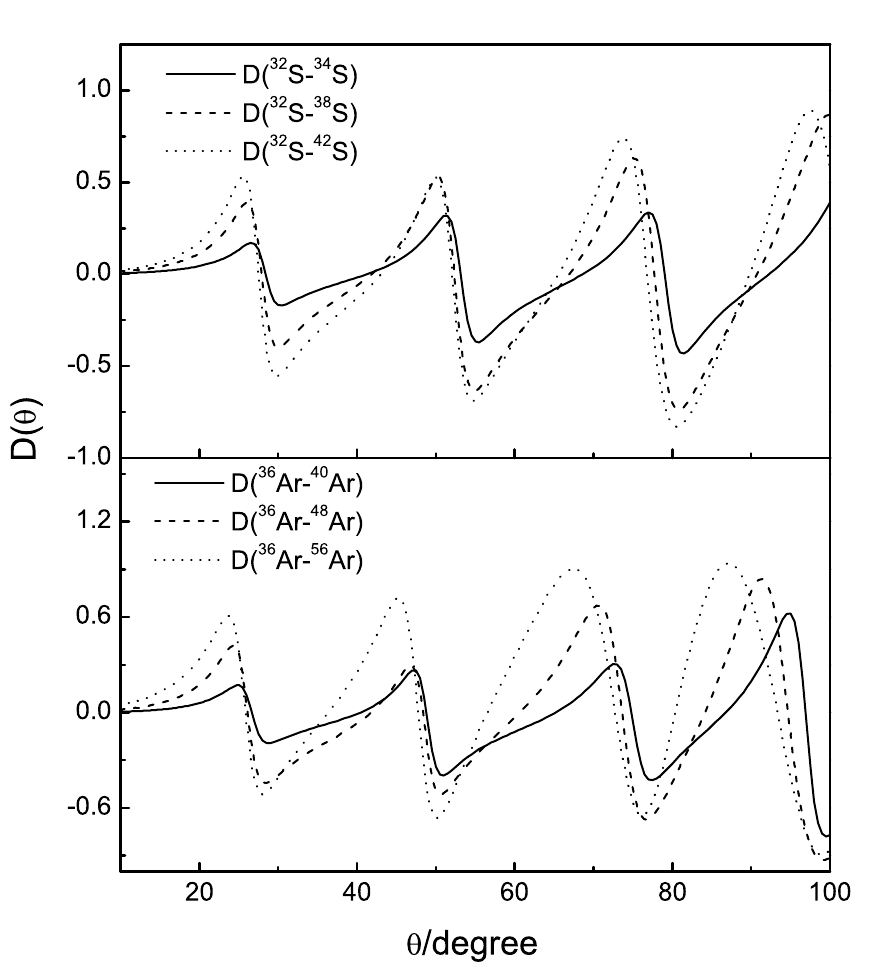}
\vspace{0cm}\caption{Plot of the differential cross section
differences $D(\theta)$. The upper panel is  between
$^{34,38,42}$S and the stable isotope $^{32}$S, and the lower
panel is between $^{40,48,56}$Ar and the stable isotope
$^{36}$Ar.}
\end{figure}
\begin{figure}[htb]
\includegraphics[width=7cm]{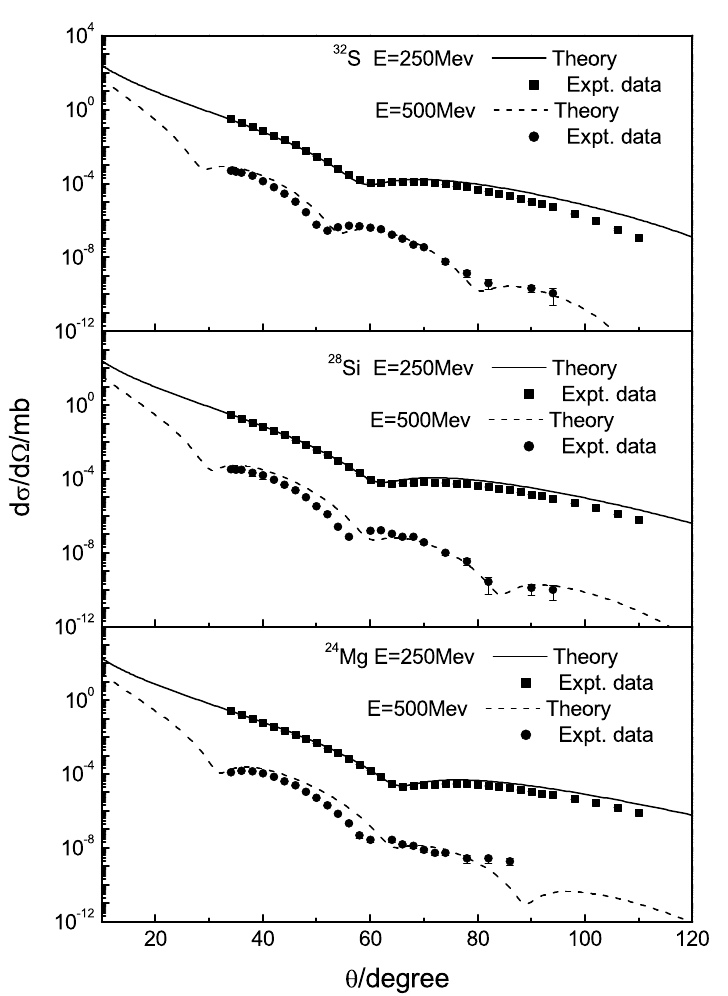}
\vspace{0cm}\caption{Comparison of the experimental differential
cross sections \cite{Sic1} with the theoretical results for the
nuclei $^{32}$S, $^{28}$Si and $^{24}$Mg.}
\end{figure}
and $^{48,56}$Ar have central depletions, while those of $^{32}$S
and $^{36}$Ar have not. It can be noted from Figure 7 that the
differential cross section differences between the isotope with
central charge density depression and the stable one without
central charge density depression are considerably large and
measurable. This also reveals that the charge density differences
shown in Figure 4 are observable and can possibly be measured with
elastic electron-nucleus scattering experiments. The level
inversion of 2$s_{1/2}$ and 1$d_{3/2}$ and central charge density
depression can further be investigated by analyzing and comparing
the experimental charge density distributions with the theoretical
results. In addition, the comparison of the theoretical results
with experimental data can be a new test of the effectiveness of
the RMF model in describing unstable nuclei.

To guarantee that the theoretical results of electron-nucleus
scattering are valid, we have tested our calculations with stable
nuclei $^{32}$S, $^{28}$Si, $^{24}$Mg with electron-nucleus
scattering experimental data available \cite{Sic1}. In the
calculations, the charge density distributions for $^{32}$S,
$^{28}$Si, $^{24}$Mg are produced by the RMF model with the NL-SH
parameter set. Figure 8 is the comparison of the calculated
results and the experimental data. It can be found from the figure
that the theoretical differential cross sections agree well with
the experimental data except some slight discrepancies in the
large scattering angle $\theta>100^{0}$ region for incident energy
$E=250$Mev. This shows that the theoretical calculations of the
electron-nucleus scattering are reliable, as well as that the RMF
model is effective and reliable to stable nuclei.

In addition to elastic electron nucleus scattering, another
possible way to investigate the 2$s_{1/2}$ and 1$d_{3/2}$ level
inversion by experiment is the pick-up or stripping nuclear
reaction experiments. The RMF model calculations show that
$^{43}$P, with a magic neutron number, may have the 2$s_{1/2}$ and
1$d_{3/2}$ level inversion. If this true, then for the ground and
first excited states of $^{43}$P the single proton should be in
the 1$d_{3/2}$ state and the 2$s_{1/2}$ state, respectively. Then,
as a result, the corresponding ground-state and
first-excited-state spin and parity of $^{43}$P should be
$I^{\pi}=(\frac{3}{2})^+$ and $I^{\pi}=(\frac{1}{2})^+$. Thus, if
we can measure the spin and parity of $^{43}$P and compare them
with the ground-state spin and parity of the stable isotope
$^{31}$P ($I^{\pi}=(\frac{1}{2})^+$), perhaps we will know if
there exists 1$d_{3/2}$ and 2$s_{1/2}$ energy level inversion in
$^{43}$P. A possible method might be to conduct pick-up or
stripping nuclear reaction and $\gamma$-decay experiments with
nuclei $^{44}$S and $^{43}$P by using the RIB technology in the
future. For instance, if we could produce $^{44}$S with the RIB
technology and bombard $^{44}$S with a mono-energetic neutron beam
to produce $^{43}$P and deuterons ($^{2}$H) or bombard $^{44}$S
with a mono-energetic triton ($^{3}$H) beam to produce $^{43}$P
and the $\alpha$-particles $^{4}$He, the excited energies, spin
and parity might be deduced from the measurements of the energy
distribution and the angular distribution of deuterons or
$\alpha$-particles and also from the angular distributions of the
photons resulting from the subsequent $\gamma$-decays of the
excited states of $^{43}$P.

\section{Summary}
In summary, we calculated the energy levels, the proton occupation
probabilities of the 1$d_{5/2}$, 2$s_{1/2}$ and 1$d_{3/2}$ states
for the even-even nuclei $^{26-40}$Mg, $^{28-46}$Si, $^{30-48}$S,
$^{32-56}$Ar, as well as the charge density distributions of these
nuclei by using the RMF model with the NL-SH parameter set. The
calculations show that the level inversion of 2$s_{1/2}$ and
1$d_{3/2}$ states may occur for the magnesium, silicon, sulphur,
and argon isotopes with more neutrons away from the stability
line, and this will consequently leads to a large measurable
central depletion to the charge density distributions for the
neutron-rich sulphur and argon isotopes. The charge density
differences between the isotopes with and without central charge
density depletion can reveal not only the level inversion of the
2$s_{1/2}$ and 1$d_{3/2}$ states but also the behavior of the
proton wave functions of both states. Electron-nucleus scattering
is an excellent way in measuring the nuclear charge density
distribution, so it is expected that the 2$s_{1/2}$ and 1$d_{3/2}$
state level inversion and the central charge depletion can
possibly investigated experimentally with electron scattering off
short-lived nuclei on the new-generation electron-nucleus collider
in the future. For comparison and reference, we also calculated
the charge form factors and differential cross section
differences. In addition to electron-nucleus scattering
experiments, if possible, we also propose to explore the possible
level inversion of the 2$s_{1/2}$ and 1$d_{3/2}$ states with
nuclear reaction $^{44}$S(n, d)$^{43}$P or $^{44}$S($^{3}$H,
$\alpha$)$^{43}$P with the RIB technology in the future.

\begin{center}
{\large Acknowledgments }
\end{center}
This work is supported by the National Natural Science Foundation
of China (Grand No.11275138, 10975072, and 10675090) and by the
Research Fund of Tianjin University of Technology and Education
(Grand No. KJYB11-3).

\end{document}